\documentstyle[12pt]{article}

\makeatletter
\def\slash#1{{\mathpalette\c@ncel{#1}}} % TeXbook, bottom of p360
\makeatother

\newcommand\beq{\begin{eqnarray}}
\newcommand\eeq{\end{eqnarray}}

\newcommand\la{\langle}
\newcommand\ra{\rangle}

\def\nslash{\slash{\mkern-1mu n}}

\def\bbar{\bar{b}}
\def\cbar{\bar{c}}

\def\shat{\hat{s}}
\def\that{\hat{t}}
\def\uhat{\hat{u}}
\def\Svec{\vec{S}}
\def\pvec{\mbox{\boldmath $p$}}
\def\kvec{\mbox{\boldmath $k$}}

\begin{document}

%%%%%%%%%%% Titlepage

\begin{titlepage}
\vskip0.5cm
\begin{center}
  {\Large \bf 
Estimate of a 
Chiral-Odd Contribution to Single Transverse-Spin Asymmetry
in Hadronic Pion Production
  \\}

\vspace{1cm}
 {\sc Y.~Kanazawa and Yuji~Koike}
\\[0.3cm]
\vspace*{0.1cm}{\it Department of Physics, Niigata University,
Ikarashi, Niigata 950--2181, Japan}
\\[1cm]

%{\em Version of \today}\\[1cm]

  \vskip1.8cm
  {\large\bf Abstract:\\[10pt]} \parbox[t]{\textwidth}{ 

We present an estimate of a chiral-odd contribution to 
the single transverse spin asymmetry
in the large-$p_T$ pion production in the nucleon-nucleon collision. 
In this contribution
the transversity distribution and the chiral-odd spin-independent
twist-3 distribution appear.  Because of the smallness of
the corresponding hard cross section, 
this term turned out to be
negligible in all kinematic regions compared with the chiral-even 
contribution.

}

\end{center}

\vskip1cm


\noindent
[Keywords: Single transverse spin asymmetry, Twist three, Chiral-odd]

  \vskip1cm 

\end{titlepage}

\setcounter{equation}{0}

Single transverse 
spin asymmetries for the pion production with large transverse momentum 
in $pp$ collision
\beq
N'(P',\Svec_\perp) + N(P) \to \pi(\ell) + X,
\label{single}
\eeq
have been 
receiving great attention\,\cite{ET}-\cite{KK}, 
in particular, after the FNAL E704 data showed 
a large asymmetry\,\cite{Adams}.  
Ongoing experiment at RHIC is expected to provide more data
on the asymmetry.  
The process probes particular quark-gluon correlations (higher twist effect)
\,\cite{ET,QS99,KK,QS91}
in the nucleon not included in the twist-2
parton densities. 
According to the 
generalized QCD factorization theorem\,\cite{CSS}, 
the polarized cross section for this process
consists of three types of twist-3 contributions:
\beq
&{\rm (A)}&\quad G_a(x_1',x_2')\otimes q_b(x)\otimes D_{c\to \pi}(z)\otimes
\hat{\sigma}_{ab\to c},\\
&{\rm (B)}&\quad
\delta q_a(x')\otimes E_b(x_1,x_2) \otimes D_{c\to \pi}(z)\otimes 
\hat{\sigma}_{ab\to c}',\\
&{\rm (C)}&\quad
\delta q_a(x')\otimes q_b(x) \otimes D^{(3)}_{c\to \pi}(z_1,z_2)\otimes
\hat{\sigma}_{ab\to c}''.
\eeq  
Here
the functions $G_a(x_1',x_2')$, $E_b(x_1,x_2)$ 
and $D^{(3)}_{c\to\pi}(z_1,z_2)$
are the twist-3 quantities representing, respectively, the
transversely polarized distribution, the unpolarized distribution, and
the fragmentation function for the pion, and
$a$, $b$ and $c$ stand for the parton's species.
Other functions are twist-2; 
$q_b(x)$ the unpolarized distribution (quark or gluon) and
$\delta q_a(x)$ the transversity distribution, etc. 
The symbol $\otimes$ denotes convolution. 
$\hat{\sigma}_{ab\to c}$ {\it etc} represents the partonic cross section
for the process
$a+b \to c + anything$ which yields large transverse momentum of
the parton $c$. 
Note that 
$\delta q_a$, $E_b$ and $D^{(3)}_{c\to\pi}$
are chiral-odd, and (B) and (C) contain two chiral-odd functions.  

Qiu and Sterman\,\cite{QS99} presented a first systematic QCD analysis
on the chiral-even contribution (A).  They showed that 
at large $x_F$, i.e., 
pion production in the forward direction with respect to the polarized 
nucleon beam  
which mainly probes large $x'$ and small $x$ region,
the cross section is dominated by the particular terms in (A)
which contain the derivatives of the {\it valence} twist-3 distribution
$G_{Fa}(x',x')$.  The main reason for this observation is 
the relation 
$|{\partial \over \partial x'}G_{Fa}(x',x')| \gg G_{Fa}(x',x')$
owing to the behavior of $G_{Fa}(x',x')
\sim (1-x')^\beta$ ($\beta >0$) at $x'\to 1$.
Keeping only those terms  
({\it valence quark-soft gluon approximation})
together with a moderate model assumption for $G_{Fa}$,
they reproduced the rising behavior of the
E704 data toward $x_F\to 1$ reasonablly well.  In a recent paper\cite{KK}, 
we extended
the analysis to one of the chiral-odd contribution ((B) term)
and presented a cross section formula in this valence qaurk-soft 
gluon approximation.  
We also discussed a possibility that this term could be a large source of the
asymmetry at large {\it negative} $x_F$ in parallel with the argument
for the chiral-even contribution.   
In this report, we shall present 
an actual estimate of the chiral-odd contribution (B)
in comparison with the chiral-even one (A). 
We will see that unlike the previous expectation
the chiral-odd one derived in \cite{KK} is negligible in all kinematic range
due to the smallness of the hard cross section, even though
the derivative brings an enhancement to 
$E_b(x,x)$.

The polarized cross section for (\ref{single}) is a function of 
three independent variables, 
$S=(P+P')^2\simeq 2P\cdot P'$, 
$x_F = 2\ell_{\parallel}/ \sqrt{S}$ ($=(T-U)/S$ below), 
and $x_T = 2\ell_{T}/ \sqrt{S}$.    
$T=(P'-\ell)^2\simeq -2P'\cdot \ell$ and 
$U=(P-\ell)^2\simeq -2P\cdot \ell$ are given in terms of these three
variables by 
$T= -S\left[ \sqrt{x_F^2 + x_T^2} - x_F\right]/2$ and 
$U= -S\left[ \sqrt{x_F^2 + x_T^2} + x_F\right]/2$.
In this convention, 
production of the pion
in the forward (backward) hemisphere in the direction of 
the polarized nucleon
corresponds to $x_F>0$ ($x_F <0$).  
Since $-1<x_F <1$, $0<x_T<1$ and $\sqrt{x_F^2 + x_T^2} < 1$,
$x_F \to 1$ corresponds to the region with $-U\sim S$ and $T\sim 0$, 
and $x_F\to -1$ corresponds to the region with $-T\sim S$ and $U\sim 0$.

In the valence quark-soft gluon approximation, the cross section 
formula for (A) and (B) terms read, respectively\,\cite{QS99,KK}
\beq
E_\pi{d^3\Delta\sigma^A(S_\perp) \over d \ell^3}
&=&{\pi M\alpha_s^2 \over S}\sum_{a,c}\int_{z_{min}}^1
{d\,z\over z^3}{D}_{c\to\pi}(z)
\int_{x_{min}'}^1 {d\,x'\over x'}
{1\over x'S + U/z}
\int_0^1 {d\,x\over x}
\delta\left(x+{x'T/z \over x'S + U/z}\right)\nonumber\\
& \times & 
\epsilon_{\ell S_\perp p n}
\left({1\over -\hat{u}}\right)\left[ -x' {\partial \over \partial x'}
G_{Fa}(x',x')\right]
\left[ 
G(x)\Delta \widehat{\sigma}_{ag\to c} +\sum_{b} q_b(x) 
\Delta \widehat{\sigma}_{ab\to c}
\right]
\label{even},\\
E_\pi{d^3\Delta\sigma^B(S_\perp) \over d \ell^3}
&=& {\pi M \alpha_s^2 \over S}\sum_{a,b,c}\int_{z_{min}}^1
{d\,z\over z^3}{D}_{c\to\pi}(z)
\int_{x_{min}}^1 {d\,x\over x}
{1\over xS + T/z}
\int_0^1 {d\,x'\over x'}
\delta\left(x'+{xU/z \over xS + T/z}\right)\nonumber\\
& \times & 
\epsilon_{\ell S_\perp p n}
\left({1\over -\hat{t}}\right)\left[ -x {\partial \over \partial x}
E_{Fb}(x,x)\right]\delta q_a(x')\delta\widehat{\sigma}_{ab\to c},
\label{odd}
\eeq
where 
$p$ and $n$ are the two light-like vectors defined from the momentum of the
unpolarized nucleon as $P=p+M^2n/2$, $p\cdot n=1$ and
$\epsilon_{\ell S_\perp p n}=
\epsilon_{\mu\nu\lambda\sigma}\ell^\mu S_\perp^{\nu} p^\lambda n^\sigma
\sim {\rm sin}\phi$ with $\phi$ the azimuthal angle between the nucleon spin
vector and the scattering plane.  
The invariants in the parton level are defined as
\beq
\shat &=&(p_a + p_b)^2 \simeq (x'P' + xP)^2 \simeq xx'S,\nonumber\\
\that &=&(p_a-p_c)^2 \simeq (x'P'-{\ell/z})^2 \simeq x'T/z,\nonumber\\
\uhat &=&(p_b-p_c)^2 \simeq (xP-{\ell/z})^2 \simeq xU/z,
\eeq
and the lower limits for the integration variables are
\beq
z_{min} &=& {-(T+U) \over S}=\sqrt{x_F^2+x_T^2},\nonumber\\
x_{min} &=& {-T/z \over S+U/z},\qquad x_{min}' = {-U/z \over S+T/z}. 
\eeq
Equation (\ref{even}) is 
derived in \cite{QS99} with the unpolarized gluon distribution $G(x)$
and the partonic cross section 
$\Delta\widehat{\sigma}_{ag\to c}$ and $\Delta\widehat{\sigma}_{ab\to c}$.  
$G_F(x',x')$ is the soft gluon component of the
twist-3 transversely polarized
distribution:
\beq
G_{Fa}(x,x)={1\over M}\epsilon_{S_\perp \sigma p n}
\int{d\lambda \over 2\pi}e^{i\lambda x}\la PS | \bar{\psi}^a(0)\nslash
\left\{ \int {d\mu\over 2\pi}g F^{\sigma\beta}(\mu n)n_\beta\right\}
\psi^a(\lambda n)|PS\ra,
\label{GF}
\eeq
where $\epsilon_{S_\perp \sigma p n}\equiv \epsilon_{\mu\sigma\nu\lambda}
S_\perp^\mu p^\nu n^\lambda$.  
Equation (\ref{odd}) is 
derived
in \cite{KK} with the partonic cross section
$\delta \hat{\sigma}_{ab\to c}$.
$E_F(x,x)$ is the soft gluon component of the
unpolarized twist-3 distrbution defined as
\beq
E_{Fa}(x,x)={-i\over 2M}
\int{d\lambda \over 2\pi}e^{i\lambda x}\la P | \bar{\psi}^a(0)\nslash
\gamma_{\perp\sigma}
\left\{ \int {d\mu\over 2\pi}g F^{\sigma\beta}(\mu n)n_\beta\right\}
\psi^a(\lambda n)|P\ra.  
\label{EF}
\eeq
The summation for the flavor indices of $G_{Fa}(x',x')$ and $E_{Fa}(x,x)$ 
is to be over
$u$- and $d$- valence quarks,   
while that for the twist-2 distributions is 
over $u$, $d$, $\bar{u}$, $\bar{d}$,
$s$, $\bar{s}$. 
$\Delta\widehat{\sigma}_{ab\to c}$ and  
$\Delta\widehat{\sigma}_{ag\to c}$ 
can be obtained from the $2\to 2$ diagrams shown in Figs. 1 and 2,
respectively.
$\delta\widehat{\sigma}_{ab\to c}$ is also obtained from diagrams in
Fig. 1 but with different spin projection from the chiral even case.
Because of the chiral-odd nature of $\delta q(x')$ and $E_F(x,x)$, they 
have to form a closed Fermion loop together, and accordingly
the contribution from the diagrams shown in Figs. 1(a) and (b)
vanishes.  Gluon contribution shown in Fig. 2 is also absent
in (\ref{odd}).

At large $|x_F|$, 
one can easily guess the relative magnitude of each diagram
shown in Fig.1.   The gluon propagators 
in Figs. 1(a), (b), (c) and (d) give rise to the factors
$1/\hat{t}^2$, $1/\hat{u}^2$, $1/\hat{s}^2$ and $1/\hat{t}\hat{u}$.
At $x_F\to 1$, $\hat{t}$ becomes very small and the 
hard cross section contribution from Fig.1(a)
becomes much larger than the others.
At $x_F \to -1$ the role of $\that$ and $\uhat$ is interchanged. 
By the same reason, the gluon contribution (Fig.2) also becomes 
very large.
Accordingly the hard cross sections for (\ref{odd})
at $x_F\to -1$ is much smaller than those
for (\ref{even}).  
For completeness we list the partonic cross sections here.
They read
\beq
\Delta\hat{\sigma}_{ab\to c}&=&{4\over 9}\left( {\shat^2 +\uhat^2\over \that^2}
\right)\left[{1\over 4} + {1\over 8}\left(1+{\uhat\over\that}\right)\right]
\delta_{ac}
+{4\over 9}\left( {\shat^2 +\that^2\over \uhat^2}
\right)\left[{1\over 4} - {7\over 8}\left(1+{\uhat\over\that}\right)\right]
\delta_{bc}\nonumber\\
& &+{-8\over 27}
\left( {\shat^2 \over\that \uhat}
\right)\left[{10\over 8} + {1\over 8}\left(1+{\uhat\over\that}\right)\right]
\delta_{ac}\delta_{bc},\nonumber\\
\Delta\hat{\sigma}_{a\bbar\to c}
&=&{4\over 9}\left( {\shat^2 +\uhat^2\over \that^2}
\right)\left[{7\over 8} + {1\over 8}\left(1+{\uhat\over\that}\right)\right]
\delta_{ac}
+{-4\over 9}\left( {\that^2 +\uhat^2\over 
\shat^2}
\right)\left[{1\over 8} + {7\over 8}\left(1+{\uhat\over\that}\right)\right]
\delta_{ab},\nonumber\\
\Delta\hat{\sigma}_{a\bbar\to \cbar}
&=&{4\over 9}\left( {\shat^2 +\that^2\over \uhat^2}
\right)\left[{7\over 8} - {1\over 4}\left(1+{\uhat\over\that}\right)\right]
\delta_{bc}
+{-4\over 9}
\left( {\that^2 +\uhat^2\over \shat^2}
\right)\left[{1\over 8} + {1\over 4}\left(1+{\uhat\over\that}\right)\right]
\delta_{ab},
\label{evenhard}
\eeq
for the chiral-even one\footnote{The cross sections listed in eq. (76b) of
Ref.\cite{QS99} is not correct, since some of the color factors in table II
of \cite{QS99} is wrong.  Numerically, however, the difference 
between (\ref{evenhard}) and eq.(76b) of \cite{QS99} is tiny.
We thank Jianwei Qiu for communication to clarify this point.
$\Delta\hat{\sigma}_{ag\to c}$ is given in eq.(76a) of \cite{QS99}.} 
and
\beq
\delta\widehat{\sigma}_{ab\to c} &=& 
\left[{10\over 27} + {1\over 27}\left(1+
{\that \over \uhat }\right)\right]\delta_{ab}\delta_{bc},\nonumber\\
\delta\widehat{\sigma}_{\bar{a}b\to c} &=& - 
{\that\uhat\over \shat^2} \left[
{1\over 9} + {7\over 9}\left( 1 + {\that\over \uhat}\right)\right]
\delta_{ab},\nonumber\\
\delta\widehat{\sigma}_{\bar{a}b\to\bar{c}} &=& -
{\that\uhat\over \shat^2} \left[
{1\over 9} + {2\over 9}\left( 1 + {\that\over \uhat}\right)\right]
\delta_{ab},
\label{oddhard}
\eeq
for the chiral-odd one\,\cite{KK}
\footnote{$\delta_{\bar{a}b}$ in the second and third equations in eq.(17)
of \cite{KK} should be changed into $\delta_{ab}$ as corrected above.}.  
Even though
the hard cross section for the chiral-odd term (B)
is small, the distribution function 
receives a large enhancement by the derivative
of $E_F(x,x)$ in (\ref{odd}).  
Employing 
a simple model assumption introduced in \cite{KK},
we shall present a numerical estimate of the asymmetry.

Our model assumption for $G_F(x,x)$ and $E_F(x,x)$ is based on 
the comparison of their explicit form (\ref{GF}) and (\ref{EF})
with the twist-2 distributions
\beq
q_a(x)&=&{1\over 2}\int{d\lambda \over 2\pi}e^{i\lambda x}\la P |
\bar{\psi}^a(0)\nslash \psi^a(\lambda n)|P\ra,
\label{GF1}\\
\delta q_a(x)&=& {i\over 2}\epsilon_{S_\perp \sigma p n}
\int{d\lambda \over 2\pi}e^{i\lambda x}\la PS |\bar{\psi}^a(0)\nslash
\gamma_\perp^\sigma \psi^a(\lambda n)|PS\ra.
\label{transversity}
\eeq
We make an ansatz 
\beq
G_{Fa}(x,x) = K_a q_a(x),
\label{QSassumption}\\
E_{Fa}(x,x) = K_a \delta q_a(x),
\label{EFq}
\eeq
with a flavor-dependent parameter $K_a$ which represents 
the effect of the gluon field with zero momentum in 
$G_F(x,x)$ and $E_F(x,x)$.  
We note that even though $E_F(x,x)$ is an unpolarized distribution,
the quarks in $E_F(x,x)$ is ``transversely polarized''. 
One may interpret that these polarized quarks in unpolarized nucleon
would be a source of the asymmetry in the forward direction 
of the unpolarized nucleon
($x_F \to -1$) in (B).

Following \cite{QS99}, we determine $K_{u,d}$ to fit the E704 data
at $x_F>0$.  Since we are only interested in the estimate of
order of magnitude, we ignore the scale dependence
of each distribution and fragmentation function.
For the unpolarized distribution $q_a(x)$,
we use the GRV LO distribution at the input scale $\mu^2=0.23$ GeV$^2$
\cite{GRV}. 
For the fragmentation function of the pion, we use the one given in 
\cite{BKK} at the input scale $\mu^2=2.0$ GeV$^2$.  
The result for the single transverse spin
asymmetry $A_T$ is shown in Fig.3 with $K_u=-K_d=0.07$.
The data is reasonablly well reproduced.  
Using these values for $K_{u,d}$, we calculated the asymmetry $A_T$
at $x_F <0$ region.  For the transversity distribution,
we use the GRSV helicity distribution $\Delta q_a(x)$
(LO, valence scenario)\cite{GRSV} assuming 
$\delta q_a(x)=\Delta q_a(x)$ at the input scale.  
The result is shown in Fig. 4.  
For comparison, we also plotted the chiral-even contribution
(\ref{even})
for $\pi^+$, although
this term is not a dominant contribution among all chiral-even ones
at $x_F<0$.
For all $\pi^{\pm,0}$, it turns out that (\ref{odd}) 
gives rise to negligible asymmetry, even smaller
than a part of the chiral-even contribution in the negative $x_F$ region.
This is because the hard cross section (\ref{oddhard}) is much smaller than 
(\ref{evenhard}).  

In the region of large $x_T$ with $x_F\sim 0$, the polarized cross section for 
(\ref{single}) probes the region of $x\sim x'\sim 1$.  Accordingly
the valence quark-soft gluon approximation again becomes valid.
This kinematic region corresponds to $T\sim U\sim -S/2$ and hence
the magnitude of each contribution in Fig.2 becomes comparable. 
We therefore plotted (\ref{even}) and (\ref{odd}) in Fig.5 
separately.  One sees that even in this region the chiral-even contribution 
is much larger than the chiral-odd one.
Experimentally, this region is almost impossible to achieve.

Another chiral-odd 
contribution (C) is yet to be analyzed for a complete 
description of the single transverse asymmetry (\ref{single}).
The calculation of the hard cross section for this term
is again reduced to the calculation of $2\to 2$ scattering diagrams
shown in Figs. 1 and 2.  For this case the diagrams 
shown in Figs. 1 (a) and (b)
and Fig. 2 also contribute, and hence may cause a large effect to the 
asymmetry.   The analysis of this term will be reported elsewhere.

A different approach to the single spin asymmetry
introduces
the so-called T-odd distribution or 
fragmentation functions with the intrinsic
transverse momentum instead of 
twist-3 distributions 
introduced here\,\cite{Sivers,Collins,Boer,BMT98}.
Corresponding to (A), (B) and (C),
this approach starts from the factoriztion assumption for the three
types of contributions to the asymmetry: 
(i) $f_{1T}^\perp (x',\pvec_\perp ')\otimes q(x)\otimes D(z)\otimes
\hat{\sigma}$,
(ii) $\delta q(x')\otimes h_1^\perp(x,\pvec_\perp)\otimes D(z)\otimes
\hat{\sigma}'$,
(iii) $\delta q(x')\otimes q(x)\otimes H_1^\perp (z,\kvec_\perp)\otimes
\hat{\sigma}''$,
where $f_{1T}^\perp$ represents distribution of
an unpolarized quark with nonzero transverse momentum inside a transversely
polarized nucleon \cite{Sivers},
$h_1^\perp$ represents distribution of a transversely polarized quark 
with nonzero tranverse monetum inside the unpolarized nucleon
\cite{Boer}, and $H_1^\perp$ represents a fragmentation function for
a transversely polarized quark
fragmenting into a pion with the transvere momentum\cite{Collins}. 
Anselmino {\it et al} fitted the E704 data for the asymmetry assuming
the above (i) or (iii) are the sole origin of the asymmetry.  
It is clear from the
present study that the second contribution (ii) is tiny
due to the smallness of the partonic cross section also in this approach. 
For the lowest order Drell-Yan single spin asymmetry, it has been shown that
the T-odd function $f_{1T}^\perp$ and $h_1^\perp$
effectively play the role of the soft gluon pole contribution 
considered here\,\cite{BMT98}.  It is interesting to
explore the connection between the present approach and those in 
\cite{Sivers,Collins,Boer,ABM}.  

To summerize, we have estimated the magnitude of one of the chiral-odd
contribution to the single transverse spin asymmetry of the hadronic 
pion production.  Due to the smallness of the partonic hard cross section,
it turns out that this term is negligible in all kinematic region.

\newpage

\large
\centerline{\bf Figure Captions}

\normalsize
\begin{enumerate}

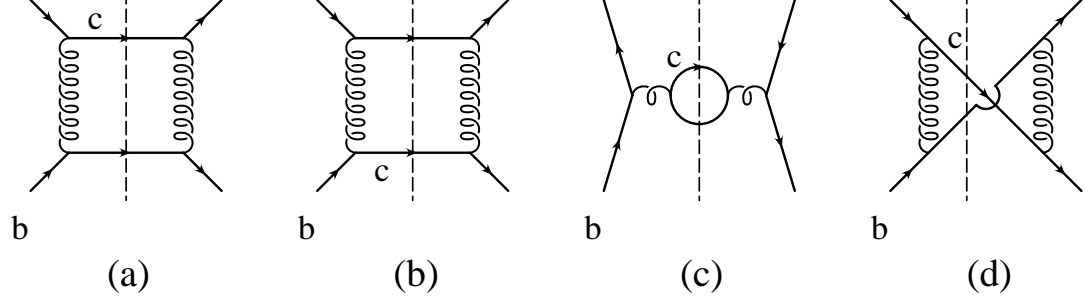
\item[{\bf Fig. 1}]
Quark-quark $2\to 2$ scattering diagrams contributing to the 
hard cross section.

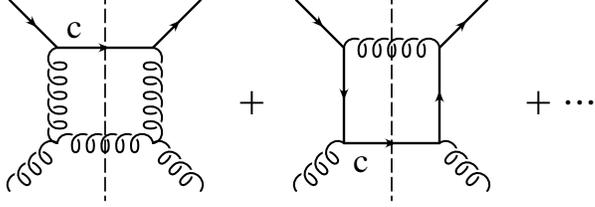
\item[{\bf Fig. 2}]
Quark-gluon $2\to 2$ scattering diagrams contributing to the 
hard cross section.

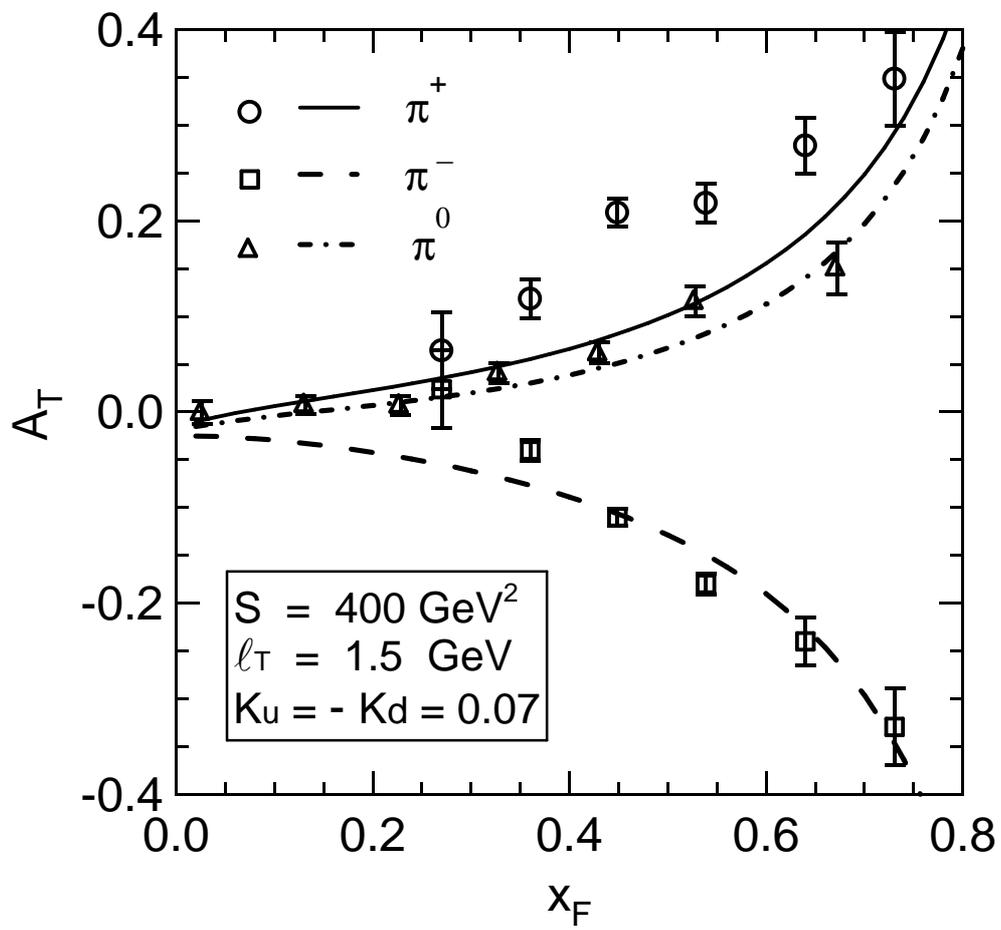
\item[{\bf Fig. 3}]
The E704 data for the
single transverse spin asymmetry $A_T$ in the inclusive pion production
in the $x_F>0$ region at $\sqrt{s}=20$ GeV\,\cite{Adams}.  
The transverse momentum of the pion is $\ell_T
=1.5$ GeV.  The curves show (\ref{even}) calculated with 
the method described in the text. 

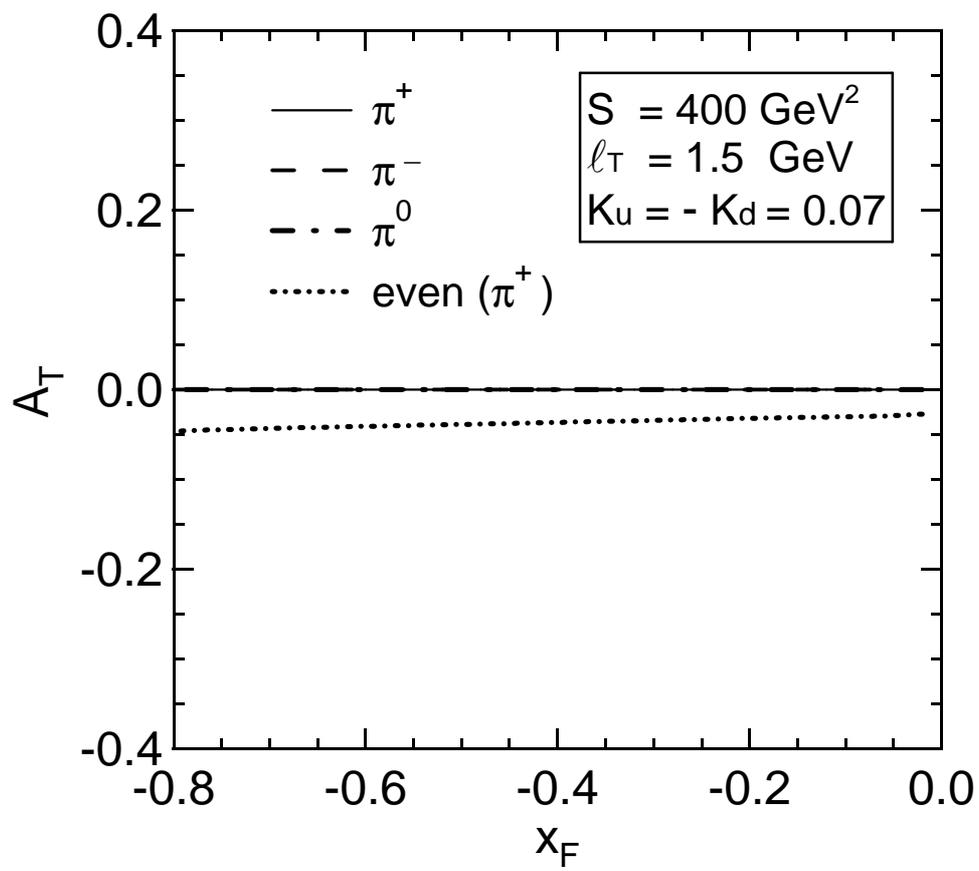
\item[{\bf Fig. 4}]
Single transverse
spin asymmetry $A_T$ in the inclusive pion production
at $x_F<0$.  The transverse momentum of the pion is calculated at $\ell_T
=1.5$ GeV.  The curves obtained from (\ref{odd}) can not be distinguished
from zero for all $\pi^{\pm,0}$.  The chiral-even contribution
(\ref{even}) is also shown by dotted line for $\pi^+$ for comparison.  

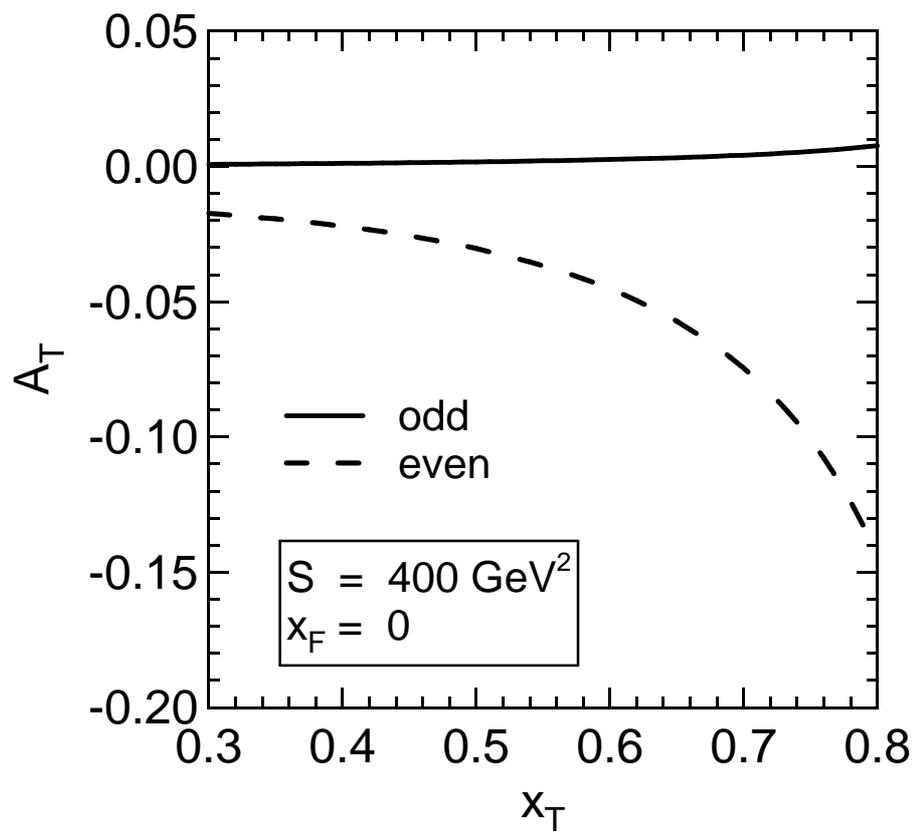
\item[{\bf Fig. 5}]
Single transverse 
spin asymmetry $A_T$ in the inclusive pion production
as a fuction of $x_T$ at $x_F=0$.  Solid and dashed lines, respectively, 
stand for the contribution from (\ref{even}) and (\ref{odd}). 

\end{enumerate}
\end{document}